\documentclass[reprint,aps,prb,onecolumn,groupedaddress,nobibnotes]{revtex4-2}
\usepackage{amssymb}
\usepackage{graphicx}
\usepackage[caption=false]{subfig} 
\usepackage{float}
\usepackage{amsmath}
\usepackage{dcolumn}
\usepackage{color}
\usepackage{graphicx}
\usepackage[pdftex,colorlinks=true,linkcolor=red,citecolor=blue]{hyperref}
\usepackage{soul}

\newcommand{\red}{\textcolor{red}}
\usepackage{setspace}
\begin{document}

\title[]{Electronic properties of kagome metal ScV$_6$Sn$_6$ using high field torque magnetometry}
\author{Keshav Shrestha$^{1\textcolor{red}{\dagger}}$}
\email{Corresponding e-mail: kshrestha@wtamu.edu}
\author{Binod Regmi$^{2}$}
\thanks{Equal contribution}
\author{Ganesh Pokharel$^{3}$}
\author{Seong-Gon Kim$^2$}
\email{Corresponding e-mail: sk162@msstate.edu}
\author{Stephen D. Wilson$^3$}
\author{David E. Graf$^{4,5}$}
\author{Birendra A. Magar$^{6}$}
\author{Cole Phillips$^{1}$}
\author{Thinh Nguyen$^{1}$}

\affiliation{$^{1}$Department of Chemistry and Physics, West Texas A$\&$M University, Canyon, Texas 79016, USA}
\affiliation{$^{2}$Department of Physics and Astronomy, Mississippi State University, Mississippi State, MS 39762, USA}
\affiliation{$^{3}$Materials Department, University of California, Santa Barbara, California 93106, USA}
\affiliation{$^{4}$Department of Physics, Florida State University, Tallahassee, FL, 32306, USA}
\affiliation{$^{5}$National High Magnetic Field Laboratory, Tallahassee, Florida 32310, USA}
 \affiliation{$^6$Department of Physics, New Mexico State University, Las Cruces, New Mexico 88003, USA}
\begin{abstract}
  This work presents electronic properties of the kagome metal ScV$_6$Sn$_6$ using de Haas-van Alphen (dHvA) oscillations and density functional theory (DFT) calculations. The torque signal with the applied fields up to 43 T shows clear dHvA oscillations with six major frequencies, five of them are below 400 T (low frequencies) and one is nearly 2800 T (high frequency). The Berry phase calculated using the Landau level fan diagram near the quantum limit is approximately $\pi$, which suggests the non-trivial band topology in ScV$_6$Sn$_6$. To explain the experimental data, we computed the electronic band structure and Fermi surface using DFT in both the pristine and charge density wave (CDW) phases. Our results confirm that the CDW phase is energetically favorable, and the Fermi surface undergoes a severe reconstruction in the CDW state. Furthermore, the angular dependence of the dHvA frequencies are consistent with the DFT calculations. The detailed electronic properties presented here are invaluable for understanding the electronic structure and CDW order in ScV$_6$Sn$_6$, as well as in other vanadium-based kagome systems.

\end{abstract}


\maketitle

Kagome materials have recently garnered enormous interest as they exhibit many interesting physical
and material properties, such as charge density wave (CDW), non-trivial topology, geometrically frustrated magnetism~\cite{Ye_2018, yin2020quantum,Kun-Review}. The atoms in kagome compounds form a quasi-2D lattice resembling Japanese basket-weaving pattern~\cite{PhysicsToday}. For example, a new vanadium-based family $A$V$_3$Sb$_5$ ($A$ = K, Rb, and Cs) form a hexagonal lattice of V atoms that are coordinated by Sb atoms~\cite{Ortiz,Ortiz1,Ortiz2}. Numerous exotic quantum phenomena, superconductivity (SC) ($T_c$ $\sim$ 0.3 - 3 K), CDW order near $T_\text{CDW}$ $\sim$ (80 - 110 K), a van Hove singularity, etc. have been discovered in this family~\cite{Yu, Chen, Wang, MKang}. Recent quantum oscillations studies~\cite{Yu1, Yin, Yang, Nakayama, Ortiz3, Luo, Fu, Zhang, Broyles,Shrestha9,Shrestha11} on $A$V$_3$Sb$_5$ have verified the non-trivial band topology and also revealed a severe reconstruction of the Fermi surface in the CDW phase.

In a search for a new kagome system, a vanadium-based $R$V$_6$Sn$_6$ ($R$ = Rare earth) (also called “166” compounds)~\cite{GaneshPRB,JeonghunPRM,PengPRL,Rosenberg,XiaoxiaoPRM,Shuting,Ganesh-PRM2022} have been discovered. Among 166 compounds, GdV$_6$Sn$_6$, HoV$_6$Sn$_6$, and YV$_6$Sn$_6$ are already known to show topological features~\cite{PengPRL, YongScienceAdv}. Arachchige \textit{et al.}~\cite{Hasitha} recently reported ScV$_6$Sn$_6$, which is the only compound in the 166 family exhibiting a CDW order below $T_\text{CDW}$ = 92 K. The $T_\text{CDW}$ transition is confirmed to be of the first-order nature ~\cite{Hasitha, Tianchen-optical}. However, no SC has been detected in ScV$_6$Sn$_6$ either under ambient or high pressure conditions up to 11 GPa~\cite{XiaoxiaoHP}. The x-ray and neutron scattering experiments~\cite{Hasitha} reveal the $\sqrt{3}\times\sqrt{3}\times 3$ structural modulation in the CDW phase of ScV$_6$Sn$_6$.  A thorough understanding of the Fermi surface is crucial to comprehend the CDW phase in ScV$_6$Sn$_6$. However, to date, there have been no reports on the electronic properties of ScV$_6$Sn$_6$ through quantum oscillations measurements.

In this work, we present electronic properties of ScV$_6$Sn$_6$ single crystal using torque magnetometry and DFT calculations. We could resolve six distinct frequencies in dHvA oscillations, five of them below 400 T and one is very high $\sim$ 2800 T. The Berry phase computed at the quantum limit is approximately $\pi$, which provides evidence for the non-trivial topology of ScV$_6$Sn$_6$. Moreover, our first principles calculations in the CDW phase are consistent with experimental results.

High-quality single crystals of ScV$_6$Sn$_6$ were synthesized using the flux-based growth technique. The side and top views of the crystal structure of ScV$_6$Sn$_6$ are displayed in Fig. 1(a). The top view clearly shows the Kagome network of V atoms in two layers. Torque measurements were performed at the National High Magnetic Field Laboratory (NHMFL) in Tallahassee, Florida, using a hybrid magnet with DC magnetic fields up to 43 T. Electronic band structure and Fermi surface were calculated using spin-polarized DFT with projector-augmented wave (PAW) potentials as implemented in Vienna Ab initio Simulation Package (VASP). The detailed sample synthesis and experimental details are presented in the Supplemental Material (SM)~\cite{Supp}. Our temperature dependent magnetic susceptibility data shows a sharp drop below 92 \red{K} due to the CDW transition that is consistent with recent reports~\cite{Hasitha,tuniz2023dynamics,cheng2023nanoscale} (Fig. S1 in SM~\cite{Supp}).

Figure 1(b) shows the magnetic torque ($\tau$) for two ScV$_6$Sn$_6$ single crystals (S1 and S2) with applied fields ($H$) up to 43 T at 0.5 K. The $\tau$ signal decreases with $H$ and shows indication of dHvA oscillations, which is clearly reflected in the background subtracted data in Fig. 1(c). The oscillations in both samples are evident and even start at relatively low magnetic fields of around 4 T. Moreover, the oscillations for both samples appear to be similar, and their frequency spectra are comparable as well (Fig. S2 in SM~\cite{Supp}).

The frequency spectra of S2 at different temperatures, obtained after performing the fast Fourier transform (FFT) of the background-subtracted quantum oscillations data, are presented in Fig. 1(d). There are six distinct frequency peaks, five of which are low frequencies ($F_\alpha$ = 77 $\pm$ 12 T, $F_\gamma$ = 155 $\pm$ 8 T, $F_\chi$ = 235 $\pm$ 10 T, $F_\delta$ = 314 $\pm$ 15 T, and $F_\epsilon$ = 394 $\pm$ 18 T) and the remaining one is high frequency $F_\beta$ = 2834 $\pm$ 30 T. Near $F_\beta$, there exists an additional peak at 2960 $\pm$ 26 T, as indicated by a dagger in the inset. We have not considered the shoulder peak as a frequency here since it is not clearly resolved at higher tilt angles ($\theta$), where $\theta$ is the angle between the normal to the sample surface and the magnetic field direction.
At higher temperatures, the amplitude of the frequencies decreases. This behavior can be explained by the Lifshitz-Kosevich (LK) formula~\cite{Shoenberg,Shrestha2}. We used the LK formula to fit the temperature-dependent FFT data and estimated the effective masses of the charge carriers to be $m^*_\alpha$ = 0.19$m_e$ and $m^*_\beta$ = 0.77$m_e$, where $m_e$ is the free electron mass, for $\alpha$ and $\beta$ pockets, respectively (Fig. S3 in SM~\cite{Supp}). These $m^*$ values are slightly higher than those observed in other 166 family compounds~\cite{HasanSdHPRL}, but they are comparable to the values reported for $A$V$_3$Sb$_5$~\cite{Yu1, Yin, Shrestha9, Broyles, Shrestha11,Shrestha12}.

The angular dependence of quantum oscillations provides the information on the shape, size, and dimensionality of the Fermi surface~\cite{Ando,Shoenberg,Shrestha3,Shrestha4}. Therefore, we carried out torque measurements at different tilt angles. The background subtracted torque data for ScV$_6$Sn$_6$ at selected $\theta$ values are shown in Fig. S4 in SM~\cite{Supp}. The low frequency oscillation appears to be present at all measured $\theta$-values up to 91$^o$ and starts at very low fields of 4 T [Fig 1(c)]. However, as shown in the inset in Fig. S4, the high-frequency oscillation appears only at high fields, as indicated by the dotted ellipse for $\theta = 7^\circ$, and it continues to show up to 56$^\circ$. To precisely determine the field value at which the high-frequency signal emerges, we randomly chose torque data at $\theta$ = 45$^\circ$ and analyzed that data at different magnetic field ranges. Our analyses revealed that the high-frequency signal emerges only above 35 T (Fig. S5 in SM~\cite{Supp}). Figures 2(a) and 2(b) represent the frequency spectra for ScV$_6$Sn$_6$ in the range of 10 T - 600 T and 1500 T - 6000 T, respectively. At higher $\theta$-values, $F_\alpha$ decreases, whereas $F_\beta$ increases, as indicated by the dashed curves. $F_\beta$ signal gets weaker at higher $\theta$-values and can be resolved only up to 56$^o$.

Figure 2(c)  shows the angular dependence of $F_\alpha$ and $F_\beta$ for both S1 and S2. Frequencies derived from S1 and S2 are comparable to one another and show the same angular variation. As expected, $F_\alpha$ is observed at all $\theta$ values up to 91$^o$, and it decreases at higher $\theta$-values. Whereas, $F_\beta$ increases at higher $\theta$-values and it nearly follows 1/cos$\theta$ behavior as indicated by the solid curve. This suggests that the Fermi surfaces corresponding to $F_\beta$ is nearly cylindrical~\cite{Ando, Ando2, Shrestha6}. Moreover, the disappearance of $F_\beta$ above 56$^o$ further supports its cylindrical shape~\cite{Ando, shrestha0, Shrestha1, Analytis}. For comparison, we have included frequencies derived from DFT calculations. As seen in the graph, the angular variations of $F_\alpha$ and $F_\beta$ are consistent with DFT results, which will be discussed in detail later. Unlike $F_\alpha$ and $F_\beta$, other frequencies ($F_\gamma$, $F_\chi$, $F_\delta$, and $F_\epsilon$) show very weak angular dependence as shown in Fig. S6 in SM~\cite{Supp}.

The analysis of our angular-dependent quantum oscillations data reveals that ScV$_6$Sn$_6$ possesses both quasi-2D and 3D Fermi surfaces. To understand the topological property, we calculated the Berry phase ($\Phi_\text{B}$) of the $\alpha$-band by constructing a Landau level (LL) fan diagram~\cite{Shoenberg,Ando,Shrestha1}. The $\Phi_\text{B}$ value is $\pi$ (or zero) for a topologically non-trivial (or trivial) system~\cite{Ando, Shrestha6,Shrestha7,Shrestha5}. The magnetic torque is given by $\overrightarrow{\tau} = V\mu_0 \overrightarrow{M} \times \overrightarrow{H} = V\mu_0 MHsin\lambda$, where $V$, $\mu_0$, and $\lambda$ represent the volume of the sample, the permeability of the free space, and the angle between $M$ and $H$ respectively. Assuming $\lambda$ = 90$^o$, the perpendicular component of the magnetization ($M_{\perp}$) with the external field can be determined from the torque data. Fig. 3(a) shows the background subtracted magnetization $\Delta M_{\perp}$ \textit{vs} 1/$H$ plot for ScV$_6$Sn$_6$. Since there are multiple frequencies present in the data, we isolated oscillations for $F_\alpha$ using a FFT bandpass filter~\cite{Shrestha9,Shrestha11}. The gray and magenta curves in Fig. 3(a) are the raw and filtered data, respectively. In the extracted oscillations, there is a single frequency of 79 T, as indicated in the inset. The LL index for the minima and maxima positions were assigned as ($N$ - $\frac{1}{4}$) and ($N$ + $\frac{1}{4}$), respectively, while constructing the LL fan diagram~\cite{Shrestha8,Shrestha11,Shoenberg,Shrestha13}.

 The LL fan diagrams for S1 and S2 are depicted in Fig. 3(b). From the linear extrapolation of the LL fan diagram data in the limit 1/$H$ $\rightarrow$ 0, we have obtained the intercept values, that is, $\Phi_\text{B}/2\pi$ = (0.43 $\pm$ 0.01) and (0.37 $\pm$ 0.01) for S1 and S2, respectively. These numbers correspond to $\Phi_\text{B}$ $\approx$ $\pi$, implying the non-trivial topology of the $\alpha$ band. Based on the linear extrapolation, we have also obtained the slopes of (80.2 $\pm$ 1.2) T and (79.5 $\pm$ 1.6) T for S1 and S2, respectively. The slope values correspond closely to $F_\alpha$ = 79 T, which provides confirmation that the linear extrapolation of the $N$ \textit{vs.} 1/$H$ data is precise in determining the intercept (and thus the $\Phi_\text{B}$ value), and that the band-pass filter retains the original dHvA oscillation signal without significant error. It is noteworthy that the use of high magnetic fields (43 T) causes the charge carriers to reach the second LL (near the quantum limit). Consequently, the $\Phi_\text{B}$ value obtained using the linear extrapolation is trustworthy and precise. Recently, S. Mizaffari, et al.~\cite{Shirin} have also observed a large anomalous Hall signal, and C. Yi, et al.~\cite{Shekhar}, have reported non-trivial topology at a lower frequency of $\sim$ 50 T using Shubnikov-de Haas oscillation measurements, which is consistent with our results.

In order to account for the experimental findings and understand the effect of the CDW order on Fermi surface, we conducted DFT calculations for both the pristine and CDW phases of ScV$_6$Sn$_6$. Electronic bands of the pristine ScV$_{6}$Sn$_{6}$ with spin-orbit coupling (SOC), obtained from our first-principles calculations, are shown in Fig. 4(a). Similar to other 166 kagome materials~\cite{GaneshPRB,Rosenberg,MandrusBandStruct}, there exist two Dirac points near the Fermi level at the K point, as indicated by the dotted circles and a flat band (the gray shaded area). Here, the inclusion of SOC results in an opening of gaps at both Dirac points, while without SOC, there is no such gap present (Fig. S7 in SM~\cite{Supp}). The two bands, namely 57 and 59, crossing the Fermi level, which are responsible for the formation of the Fermi surface of pristine ScV$_{6}$Sn$_{6}$ (see Fig. S8 in SM~\cite{Supp}). The Fermi surface exhibits more 3D FS character (band 57) containing necks in its belly, suggesting that the system is strongly bonded along the out-of-plane direction, whereas the band 59 forms the small electron pockets near the Brillouin zone boundary. These Fermi surface sheets are consistent with previous study~\cite{Tianchen-optical}.

 Recently, H. Arachchige, \textit{et al.}~\cite{Hasitha} carried out the x-ray and neutron scattering experiments and reported the presence of $\sqrt{3}\times\sqrt{3}\times 3$ structural modulation in the CDW phase of ScV$_6$Sn$_6$. Therefore, we conducted our investigation of the CDW phase using this specific configuration. Our DFT results suggest that the CDW phase is more energetically favorable than the pristine phase by 5.4 meV/unit-cell, which is in close agreement with the previously reported value of 5.8 meV/unit-cell~\cite{Tan}. Furthermore, in order to explore the lattice instability, we have calculated the phonon spectrum of ScV$_6$Sn$_6$ in both the pristine and CDW phases, as shown in Fig. S9 in the SM~\cite{Supp}. The phonon dispersion of the pristine phase shows imaginary phonon modes, especially along the \textit{A-L-H} direction, indicating that this phase is dynamically unstable in that region of the Brillouin zone, suggesting that the material could undergo structural distortions or phase transitions. However, the absence of imaginary frequencies in the CDW phase confirms the dynamical stability of the $\sqrt{3}\times\sqrt{3}\times 3$ CDW phase. Furthermore, our calculations of the density of states (DOS) in the pristine phase [Fig. S10 in SM~\cite{Supp}] indicate that the electronic states are dominated by Vanadium \textit{d} orbitals near the Fermi level. However, we observe only a limited reduction in DOS at the Fermi level in the CDW phase. This is because CDW distortion mainly from the distortion of Sc and Sn1, as observed recently in~\cite{pokharel}, but they provide nominal contribution to the DOS of ScV$_6$Sn$_6$. Our DOS results are consistent with recent studies~\cite{Saizheng} on ScV$_6$Sn$_6$.

We conducted further investigations on the impact of CDW and crystallographic distortion on the Fermi surface. We observed a substantial distortion ($\sim$ 0.2 \AA) of the Sc and Sn$_1$ atoms along the c-axis, as illustrated in Fig. S11 in SM~\cite{Supp}. This lattice distortion results in a significant reconstruction of electronic bands and Fermi surface sheets. The unfolded electronic band structure of ScV$_6$Sn$_6$ in the CDW phase is presented in Fig. 4(b). As expected, all features observed in the pristine phase are also preserved in the CDW state too. There are additional band features, especially along $\Gamma$-A direction, which is associated with the structural distortion in the CDW phase. This effective band structure is consistent with the previous studies~\cite{hu2023kagome}. Figure 4(c) show the Fermi surfaces of ScV$_6$Sn$_6$ in the CDW phase. The Fermi surfaces are composed of contributions from four bands, namely 511, 513, 515, and 517. The Fermi surfaces of bands 511, 513, and 515 resemble 2D cylinders containing hole pockets, while that for band 517 forms small electron pockets near the Brillouin zone boundary. Additionally, all of these Fermi surfaces feature small prismatic shape-like pockets at the center.

To compare with the experimental data, we measured the area of each energy isosurface forming the Fermi surface and then computed the oscillatory frequencies using the Onsager relation~\cite{Onsager_note}. The calculated angular dependence of frequencies from different bands is plotted in Fig. 2 (c) together with experimental data. As observed in the figure, while frequencies derived from both bands 511 and 513 describe the behavior of $F_\alpha$, band 511 provides a better match in terms of the value of $F_\alpha$. This implies that $F_\alpha$ originates from the prismatic shape-like pocket of the Fermi surface of band 511. We notice that $F_\alpha$ deviates slightly above 60$^\circ$. This could be due to the fact that at higher tilt angles, the dHvA oscillations get weaker, and frequency peaks are not as well-defined as at lower angles. The frequencies coming from all three bands, 511, 513, and 515, exhibit an upward trend with increasing $\theta$, similar to that of $F_\beta$. Among these frequencies, $F_\beta$ aligns with the one derived from band 513, confirming its origin from the outer cylindrical shape of the Fermi surface in band 513. Nearly 1/cos$\theta$ behavior of $F_\beta$ further confirms its origin from the cylindrical shape-like Fermi surface. As seen in Fig. 2(c), the frequency of band 517 is $\sim$ 60 T at 0$^\circ$ and increases at higher $\theta$ values. However, we have not observed any corresponding frequency peak from this band in the experiment. Furthermore, we shifted the Fermi level downward by 10 meV during DFT calculations. This adjustment, though \textit{ad hoc}, accounts for the doping effect, considering the ambiguity in the experimental Fermi level~\cite{Niraj, Shrestha8,Shrestha10}.

To summarize, we investigated the Fermiology of ScV$_6$Sn$_6$ using torque magnetometry and DFT calculations. Our findings reveal clear dHvA oscillations in the torque signal, with six distinct frequencies, five of which are below 400 T and one frequency near 2800 T. The use of NHMFL$^s$ high field facility up to 43 T is crucial for calculating the Berry phase near the quantum limit and determining the non-trivial topological feature of ScV$_6$Sn$_6$. Additionally, this high field enables the observation of the high frequency signal near 2800 T, which appears only above 35 T of the applied field. For further investigation, we also carried out the electronic band structure, Fermi surface, and phonon calculations in both pristine and CDW phases using DFT. There exist two Dirac points near the Fermi level at the K point and a flat band along $\Gamma$-M-K-$\Gamma$ direction. The Fermi surface undergoes a severe reconstruction in the CDW phase and four bands contribute to the Fermi surface. We found that the $\sqrt{3}\times\sqrt{3}\times 3$ CDW phase in ScV$_6$Sn$_6$ is energetically favorable, and our phonon calculations confirm its stability. Angular dependence of dHvA oscillations is consistent with those calculated by the DFT calculations. The detailed electronic properties of ScV$_6$Sn$_6$ presented in this study are novel and crucial for understanding the CDW order and non-trivial topology not only in ScV$_6$Sn$_6$ but also in other vanadium-based kagome families. \\

\section*{acknowledgements}
The work at the West Texas A$\&$M University is supported by the Killgore Faculty Research program, the KRC Undergraduate and Graduate Student Research Grants, and the Welch Foundation (Grant No. AE-0025). G.P. and S.W.D. gratefully acknowledge support via the UC Santa Barbara NSF Quantum Foundry funded via the Q-AMASE-i program under award DMR-1906325.
A portion of this work was performed at the National High Magnetic Field Laboratory, which is supported by National Science Foundation Cooperative Agreement No. DMR-1644779 and the State of Florida.

\bibliography{bibliography}

\begin{figure}[H]
  \centering
  \includegraphics[width=1.0\linewidth]{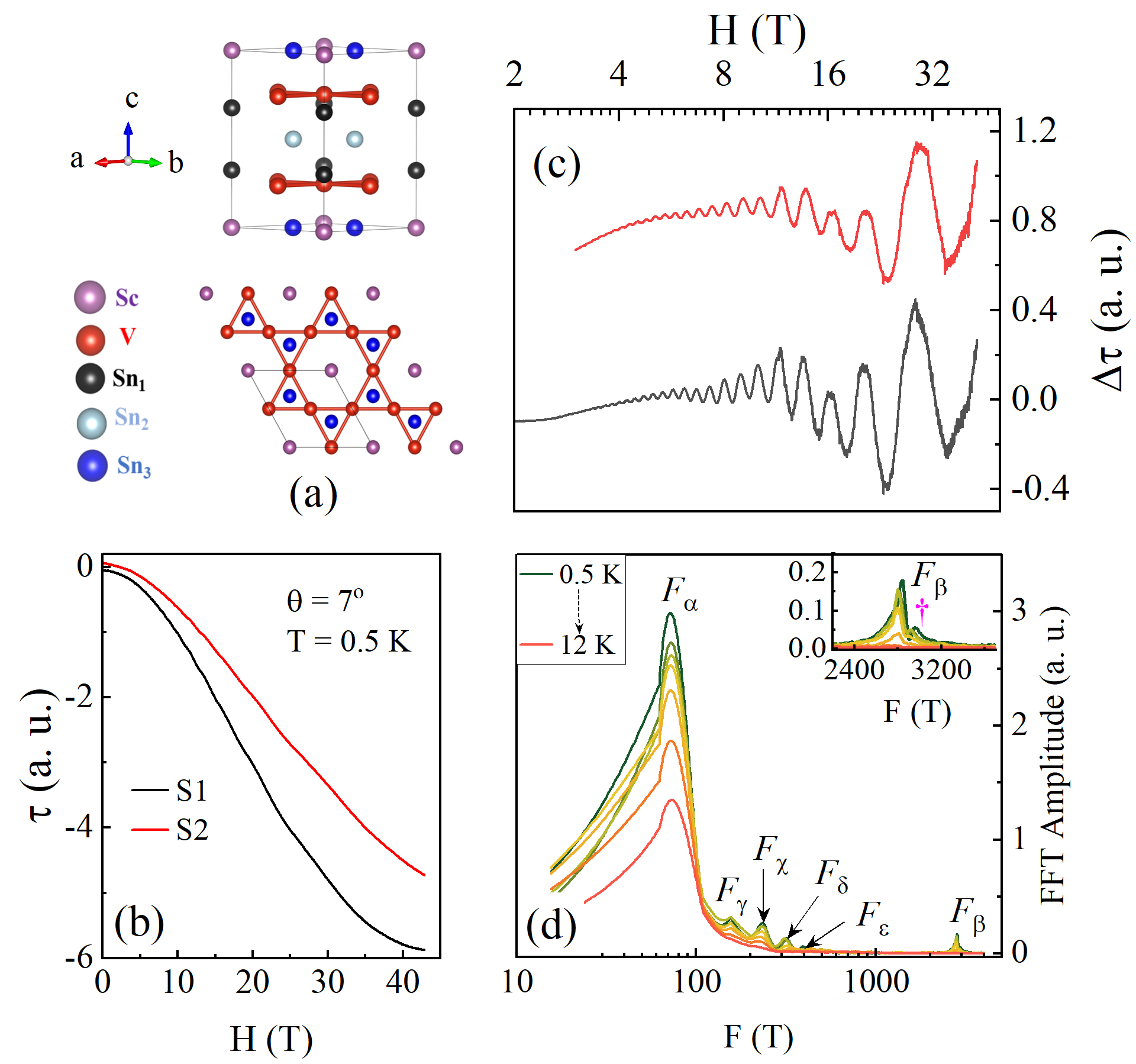}
  \caption{\doublespacing(a) A unit cell (upper panel) and top view (bottom panel) of ScV$_6$Sn$_6$, showing the kagome network consists of V-atoms. (b) Magnetic torque $\tau$ of two ScV$_6$Sn$_6$ single crystals (S1 and S2) up to 43 T at $\theta$ = 7$^o$ and $T$ = 0.5 K. (c) Background-subtracted torque $\Delta\tau$ for data shown in (b). There are clear dHvA oscillations above 4 T. (d) The FFT spectra for S2 at different temperatures. Inset: magnified view of FFT data from 2200 to 3700 T. The x-axes in (c) and (d) are in log scales for better visibility of data.}\label{Torque}
\end{figure}

\begin{figure}
  \centering
  \includegraphics[width=1.0\linewidth]{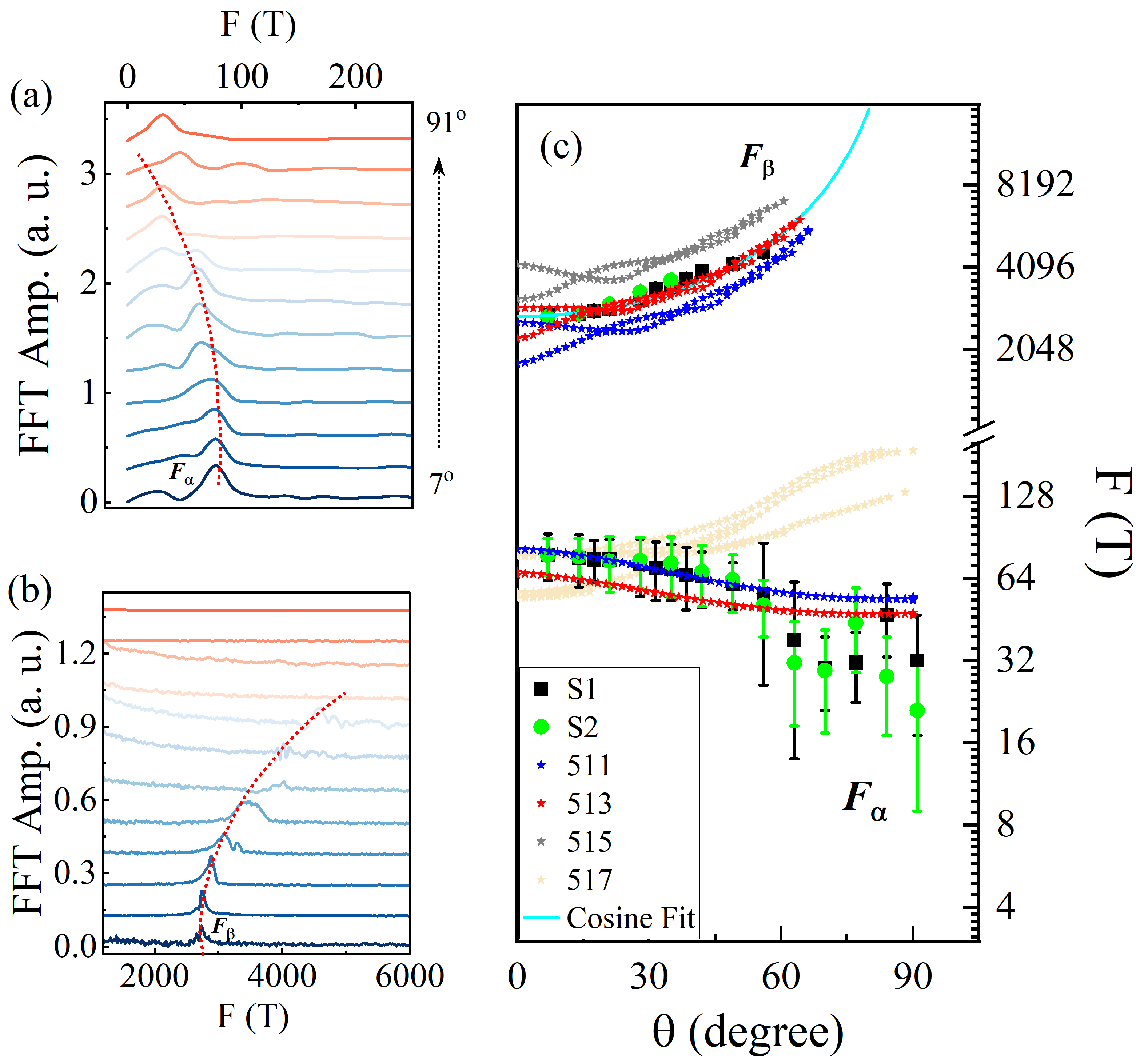}
\caption{\doublespacing Angular dependence of dHvA frequencies in the range of (a) 10 T - 600 T and (b) 2000 T - 6000 T. Both $F_\alpha$ and $F_\beta$ shift while changing $\theta$, as guided by the dotted curves. The curves in (a)-(\red{b}) are shifted vertically for clarity. (c) Angular dependence of $F_\alpha$ and $F_\beta$ for S1 (squares) and S2 (circles). $F_\alpha$ decreases at higher $\theta$ values and clearly resolved up to 91$^o$, whereas $F_\beta$ increases monotonically and disappears above 56$^o$. The error bar for each frequency data point is determined as half the width at half maximum of the respective peak in the FFT spectrum. The solid curve represents the 1/cos$\theta$ behavior of $F_\beta$. Frequencies derived from the DFT calculations are shown as dotted lines. The Fermi level in the DFT calculations was shifted downward by 10 meV for a better comparison with the experiments.}\label{Angle}
\end{figure}

 \begin{figure*}
  \centering
  \includegraphics[width=0.8\linewidth]{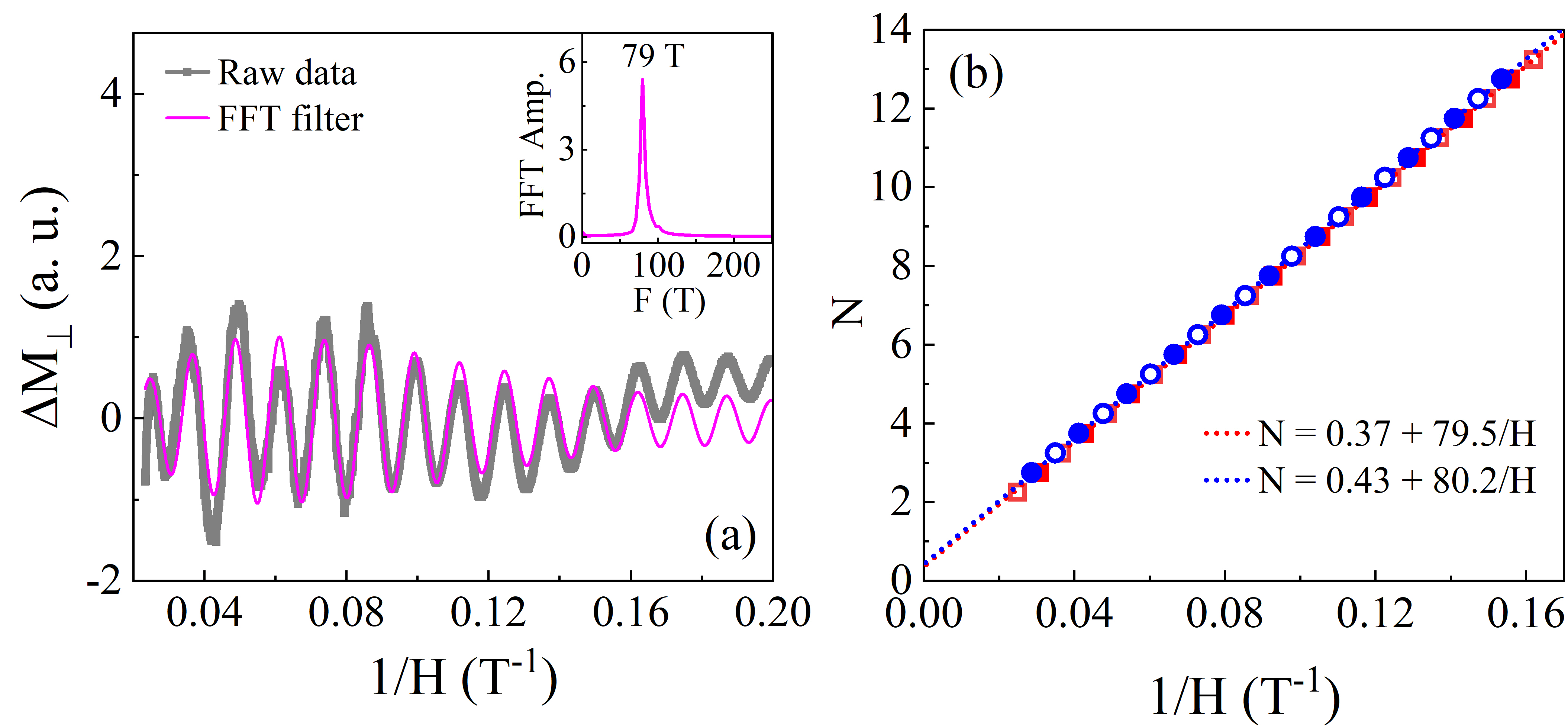}
  \caption{\doublespacing Landau level (LL) fan diagram. (a) Separation of the dHvA oscillations corresponding to $F_\alpha$ using the band-pass filter of (70 - 100 T). The grey and magenta curves represent the raw and filtered data, respectively. Inset: FFT of the processed dHvA oscillations. A single FFT peak at 79 T confirms the presence of quantum oscillations only for $F_\alpha$. (b) LL fan diagram for $F_\alpha$ for both S1 (squares) and S2 (circles). Minima (solid symbols) and maxima (open symbols) of oscillations were assigned to ($N$ - 1/4) and ($N$ + 1/4), respectively, for constructing the LL fan diagram. The dotted lines are linear extrapolations of data in the limit 1/H$\rightarrow$ 0.}\label{Berry Phase}
\end{figure*}

 \begin{figure}
  \centering
  \includegraphics[width=0.6\linewidth]{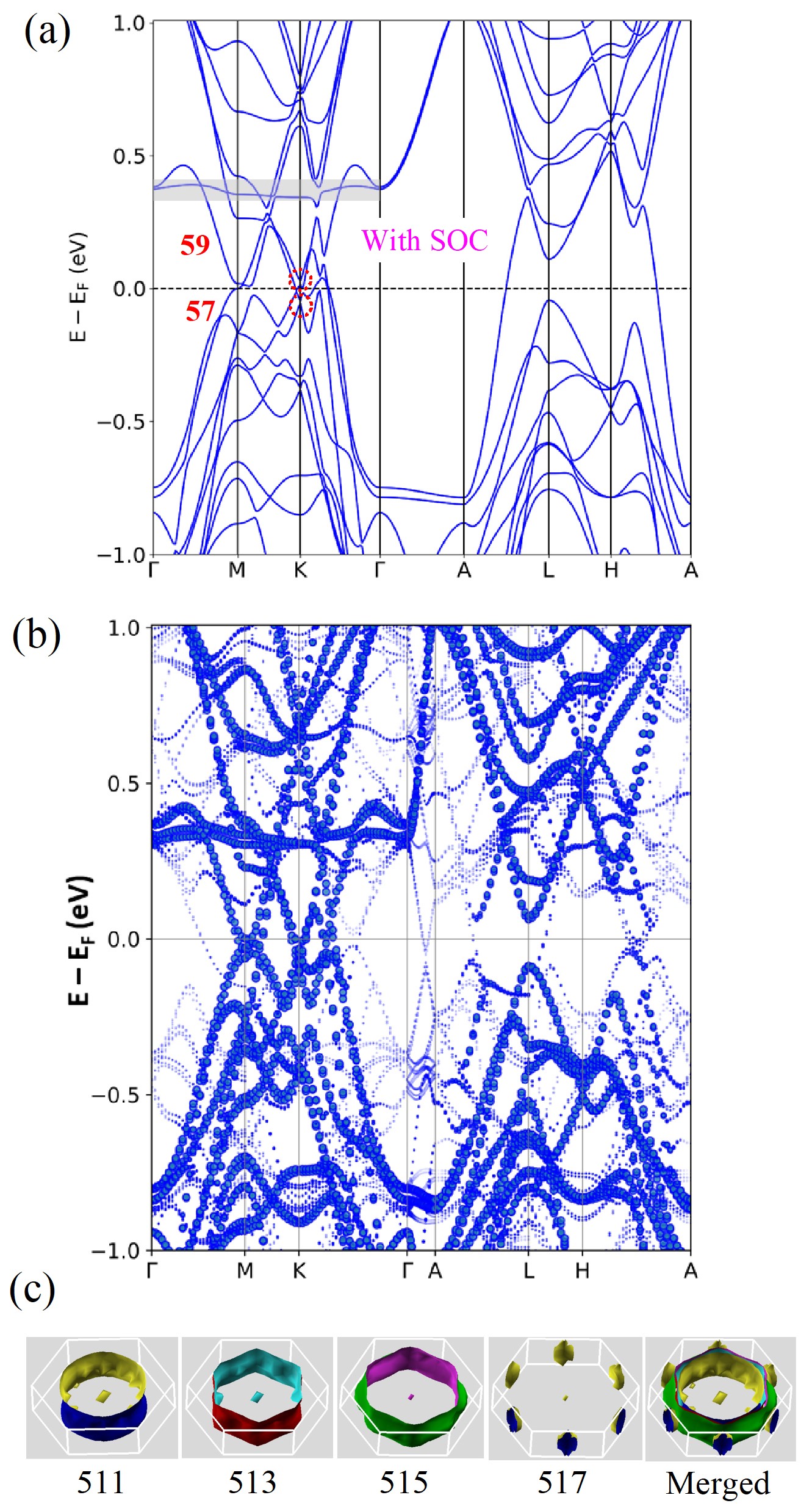}
  \caption{\doublespacing Band structure and Fermi surface. (a) Electronic band structure of pristine ScV$_6$Sn$_6$ with SOC. The flat band is denoted by the gray area, and the Dirac points near the Fermi level at the K point are indicated by dotted circles. (b) Unfolded band structure in the $\sqrt{3}\times\sqrt{3}\times 3$ CDW phase of ScV$_6$Sn$_6$. Bands of higher weight corresponds to bands from the pristine state. The pristine band features are preserved in the CDW state. (c) Band resolved Fermi surfaces of ScV$_6$Sn$_6$ in the $\sqrt{3}\times\sqrt{3}\times 3$ CDW phase.}\label{FS}
\end{figure}

\end{document}